\documentclass{JHEP3}
\usepackage{amssymb}

\def \be {\begin{equation}}
\def \ee {\end{equation}}
\def \bea {\begin{eqnarray}}
\def \eea {\end{eqnarray}}
\def \nn {\nonumber}

\def \a {\alpha}
\def \b {\beta}
\def \g {\gamma}
\def \G {\Gamma}
\def \d {\delta}

\def \m {\mu}
\def \n {\nu}
\def \k {\kappa}

\def \s {\sigma}
\def \r {\rho}
\def \o {\omega}
\def \O {\Omega}
\def \th {\theta}
\def \Th {\Theta}

\def \t {\tau}
\def \dag {\dagger}
\def \p {\partial}

\def\bd{\begin{document}}
\def\ed{\end{document}}
\def\nn{\nonumber}
\def\bea{\begin{eqnarray}}
\def\eea{\end{eqnarray}}
\let\bm=\bibitem
\let\la=\label

\def\N{{\cal N}}
\def\sst{\scriptscriptstyle}
\def\thetabar{\bar\theta}
\def\Tr{{\rm Tr}}
\def\one{\mbox{1 \kern-.59em {\rm l}}}

%

\def\a{\alpha}      \def\da{{\dot\alpha}}
\def\b{\beta}       \def\db{{\dot\beta}}
\def\c{\gamma}  \def\C{\Gamma}  \def\cdt{\dot\gamma}
\def\d{\delta}  \def\D{\Delta}  \def\ddt{\dot\delta}
\def\e{\epsilon}        \def\vare{\varepsilon}
\def\f{\phi}    \def\F{\Phi}    \def\vvf{\f}
\def\h{\eta}
\def\k{\kappa}
\def\l{\lambda} \def\L{\Lambda}
\def\m{\mu} \def\n{\nu}
\def\o{\omega}
\def\P{\Pi}
\def\r{\rho}
\def\s{\sigma}  \def\S{\Sigma}
\def\t{\tau}
\def\th{\theta} \def\Th{\Theta} \def\vth{\vartheta}
\def\X{\Xeta}
\def\z{\zeta}
\def\w{\wedge}
\def\u{\underline}
\def\hs{\hspace}


\def\cA{{\cal A}} \def\cB{{\cal B}} \def\cC{{\cal C}}
\def\cD{{\cal D}} \def\cE{{\cal E}} \def\cF{{\cal F}}
\def\cG{{\cal G}} \def\cH{{\cal H}} \def\cI{{\cal I}}
\def\cJ{{\cal J}} \def\cK{{\cal K}} \def\cL{{\cal L}}
\def\cM{{\cal M}} \def\cN{{\cal N}} \def\cO{{\cal O}}
\def\cP{{\cal P}} \def\cQ{{\cal Q}} \def\cR{{\cal R}}
\def\cS{{\cal S}} \def\cT{{\cal T}} \def\cU{{\cal U}}
\def\cV{{\cal V}} \def\cW{{\cal W}} \def\cX{{\cal X}}
\def\cY{{\cal Y}} \def\cZ{{\cal Z}}


\def\ua{\underline{\alpha}} \def\ubb{\underline{\beta}}
\def\ug{\underline{\gamma}}
\def\ub{\underline{\phantom{\alpha}}\!\!\!\beta}
\def\uc{\underline{\phantom{\alpha}}\!\!\!\gamma}
\def\um{\underline{\mu}} \def\un{\underline{\nu}}
\def\ud{\underline\delta}
\def\ue{\underline\epsilon}
\def\una{\underline a}\def\unA{\underline A}
\def\unb{\underline b}\def\unB{\underline B}
\def\unc{\underline c}\def\unC{\underline C}
\def\und{\underline d}\def\unD{\underline D}
\def\une{\underline e}\def\unE{\underline E}
\def\unf{\underline{\phantom{e}}\!\!\!\! f}\def\unF{\underline F}
\def\unm{\underline m}\def\unM{\underline M}
\def\unn{\underline n}\def\unN{\underline N}
\def\unp{\underline{\phantom{a}}\!\!\! p}\def\unP{\underline P}
\def\unq{\underline{\phantom{a}}\!\!\! q}
\def\unQ{\underline{\phantom{A}}\!\!\!\! Q}
\def\unH{\underline{H}}
\def\ul{\underline}

\def\As {{A \hspace{-6.4pt} \slash}\;}
\def\bs {{b \hspace{-6.4pt} \slash}\;}
\def\Ds {{D \hspace{-6.4pt} \slash}\;}
\def\ds {{\del \hspace{-6.4pt} \slash}\;}
\def\ss {{\s \hspace{-6.4pt} \slash}\;}
\def\ks {{ k \hspace{-6.4pt} \slash}\;}
\def\ps {{p \hspace{-6.4pt} \slash}\;}
\def\pas {{{p_1} \hspace{-6.4pt} \slash}\;}
\def\pbs {{{p_2} \hspace{-6.4pt} \slash}\;}


\def\Fh{\hat{F}}
\def\Vh{\hat{V}}
\def\Xh{\hat{X}}
\def\ah{\hat{a}}
\def\xh{\hat{x}}
\def\yh{\hat{y}}
\def\ph{\hat{p}}
\def\xih{\hat{\xi}}

\def\psit{\tilde{\psi}}
\def\Psit{\tilde{\Psi}}
\def\tht{\tilde{\th}}

\def\At{\tilde{A}}
\def\Qt{\tilde{Q}}
\def\Rt{\tilde{R}}
\def\Nt{\tilde{N}}

\def\at{\tilde{a}}
\def\st{\tilde{s}}
\def\ft{\tilde{f}}
\def\pt{\tilde{p}}
\def\qt{\tilde{q}}
\def\vt{\tilde{v}}
\def\nt{\tilde{n}}


\def\delb{\bar{\partial}}
\def\bz{\bar{z}}
\def\bD{\bar{D}}
\def\bB{\bar{B}}


\def\bk{{\bf k}}
\def\bl{{\bf l}}
\def\bp{{\bf p}}
\def\bq{{\bf q}}
\def\br{{\bf r}}
\def\bx{{\bf x}}
\def\by{{\bf y}}
\def\bR{{\bf R}}
\def\bV{{\bf V}}


\def\d{\delta}\def\D{\Delta}\def\ddt{\dot\delta}

\def\p{\partial} \def\del{\partial}
\def\xx{\times}
\def\uno{\mbox{1 \kern-.59em {\rm l}}}

\def\trp{^{\top}}
\def\inv{^{-1}}
\def\dag{{^{\dagger}}}
\def\pr{\prime}

\def\rar{\rightarrow}
\def\lar{\leftarrow}
\def\lrar{\leftrightarrow}

\def\cw{{\cal W}}
\def\cz{{\cal Z}}
\def\tcm{\tilde{\cal M}}
\def\sgn{{\rm sgn}}
\def\sd {d^{4|4}}
\def\lan{\langle}
\def\ran{\rangle}

\title{Real-time correlators in warped AdS/CFT correspondence}
\author{Bin Chen, Bo Ning and Zhi-bo Xu\\
Department of Physics,\\
and State Key Laboratory of Nuclear Physics and Technology,\\
Peking University,\\
Beijing 100871, P.R. China\\
\email{bchen01,ningbo,xuzhibo@pku.edu.cn}}


\date{\today}

\abstract{We study real-time correlators in the warped AdS/CFT
correspondence. We apply the prescription used in the usual
AdS/CFT correspondence and obtain the retarded Green's functions
for the scalar and vector fields in the spacelike warped and the
null warped black hole backgrounds. We find that the retarded
Green's functions and the cross sections are well consistent with
the predictions from dual CFT. Our results not only support
strongly the conjectured warped AdS/CFT correspondence, but also
show that the usual relativistic AdS/CFT prescription of obtaining
the real-time correlators remain effective in more general
backgrounds with anisotropic conformal infinity.

}
\newpage
\bd

\section{Introduction}

During the past few years, the AdS/CFT correspondence\cite{AdSCFT}
has been widely applied to study the physics in various many-body
strong coupling systems, ranging from quark-gluon-plasma in RHIC,
superfluid, ultra-cold atoms to superconductor. From the
computations in the gravity side, one may obtain qualitative or
even half-quantitative information on some strong coupling issues
in these systems, which otherwise is difficult, if not completely
impossible to get with the traditional methods. In particular, the
AdS/CFT correspondence has the advantage to deal with real-time
process, such as the transport behaviors of the systems. This
usually requires the calculation of real-time correlators of
composite operators of boundary quantum field theory from dual
gravity.

Compared to its Euclidean counterpart, the calculation of
real-time correlators is much subtler. For the zero-temperature
quantum field theory, real-time correlators could be obtained via
analytically continuation of Euclidean Green's
functions\cite{Witten98}. For the finite temperature case, the
dual geometry involves the black holes\cite{Witten:1998zw}, which
requires choosing appropriate boundary condition at the horizons
of the black holes. This is the first subtle point. Different
Green's functions correspond to different boundary conditions at
the horizon. It is now agreed that the retarded Green's function
corresponds to the ingoing boundary condition, while the advanced
Green's function corresponds to the outgoing one. However, even
after fixing the boundary condition, one cannot obtain the Green's
function by naively using the prescription in Euclidean version of
AdS/CFT correspondence. In \cite{Son05}, a simple prescription was
proposed to compute real-time correlators from gravity. This
prescription has been instrumental to the study of strongly
interacting system at finite temperature during the past few
years. It has also been justified from different points of view in
\cite{Herzog:2002pc,Marolf:2004fy,Gubser:2008sz,Skenderis:2008dg,Iqbal:2008by,Iqbal:2009fd}.
In particular, it was observed that the prescription proposed in
\cite{Son05} could be recast in terms of the boundary values of
the canonical conjugate momentum of the bulk fields by treating
the AdS radial direction as ``time" direction\cite{Gubser:2008sz}.
And furthermore, this reformulation was shown to be able to follow
directly from the analytic continuation of Eucliean AdS/CFT
correspondence\cite{Iqbal:2009fd}.

In this paper, we would like to apply the prescription to
calculate real-time Green's functions in the warped AdS/CFT
correspondence. The warped AdS/CFT correspondence
conjectures\cite{Andy08} that $v>1$ quantum topological massive
gravity in three dimension is holographically dual to a
two-dimensional conformal field theory with central charges
$(c_L,c_R)$ \be\label{central}
  c_L=\frac{l}{G}\frac{4v}{v^2+3}, \hspace{5ex}
  c_R=\frac{l}{G}\frac{5v^2+3}{v(v^2+3)}.
  \ee
More precisely, suitable asymptotically spacelike stretched
AdS$_3$ boundary condition on the gravity configurations needs to
be imposed in this case. Moreover, there is another
conjecture\cite{ChenXu2} stating that even at $v=1$ quantum
topological massive gravity with null warped AdS$_3$ boundary
could be dual to a two-dimensional CFT with central charges
(\ref{central}).
 These two conjectures are somehow intriguing in the
sense that the warped $AdS_3$ spacetimes have  very different
conformal boundaries from the one of $AdS_3$. Therefore, the naive
expectation that the holographic CFT resides on the asymptotic
boundary seems not true any more. Nevertheless, the discussion on
 black hole thermodynamics   and the study of the quasi-normal modes of the
warped black holes\cite{ChenXu09,ChenXu2} support the conjecture.
Moreover, for the spacelike stretched case, the physical
asymptotic boundary condition has been discussed in
\cite{Compere:2008cv,Blagojevic:2009ek}, from which the central
charges were derived from the Virasoro algebra and current
algebra. And the stability of stretched spacelike AdS$_3$ was
investigated in \cite{Anninos:2009zi,Compere:2009zj}, which paves
the ground of the correspondence.

The motivation of our study on real-time correlators in warped
AdS/CFT correspondence is two-fold. On one hand, the prescription
to get real-time correlators has been focused on AdS backgrounds,
with the dual field theory resides at the conformal boundary. It
would be very valuable to apply the prescription to more general
backgrounds which have nontrivial conformal boundaries and check
its effectiveness, especially considering the recent trend in
AdS/CMT. On the other hand, in  the warped AdS$_3$ black hole
backgrounds, since the wave functions of the perturbations could
be calculated exactly, the retarded Green's functions could be
obtained analytically. As the Green's functions in two-dimensional
CFT is well determined by the conformal invariance, this gives
another nontrivial check of the warped AdS/CFT correspondence. We
will show that real-time correlators from gravity are well
consistent with the CFT predictions. And from the retarded
correlators, we can read the quasi-normal modes and also the cross
section of the fields scattering the black hole, both of which are
in good agreement with the CFT prediction.

The remaining parts of this paper are organized as follows. In
section 2, we give a brief introduction to the prescription for
computing real-time Green's functions from the gravity side in
AdS/CFT correspondence. In section 3, we outline the various kinds
of Green's functions in two-dimensional conformal field theory. In
section 4, we discuss the retarded Green's functions for the
scalar and vector perturbations in the spacelike stretched AdS$_3$
black holes. In section 5, we turn to the null warped case. In
section 6, we investigate the extremal spacelike stretched  warped
black hole, which has different eigenfunctions from the
non-extremal ones. We end with some discussions in section 7.

\section{Retarded Green's functions from gravity: prescription}

In this section, we give a brief review of the prescription for
computing two-point Green's functions from gravity. The detailed
discussion could be found in
\cite{Son05,{Gubser:2008sz},{Iqbal:2009fd}}.

In Eulidean space, the core relation of AdS/CFT correspondence
is\cite{Witten98}
 \be\label{core}
 \langle e^{\int_{\p M}\phi_0(x){\cal O}(x)}\rangle_{\mbox{\tiny QFT}} =
 e^{-S_{\mbox{\tiny grav.}}[\phi_0]},
 \ee
 where ${\cal O}$ is a boundary CFT operator coupled to the
 bulk field $\phi$. $S_{\mbox{\tiny grav.}}[\phi_0]$ is the bulk action
 for $\phi$ evaluated at the classical solution $\phi_E$ which is
 regular in the interior and asymptotical to $\phi_0$ at the boundary.
 The two-point Green's function $<{\cal O}(x){\cal O}(0)>$ could be
 obtained from taking the second functional derivative of
 $S_{\mbox{\tiny grav.}}$ with respect to $\phi_0$. However there is an
 equivalent way to compute the Green's function. From the above
 relation (\ref{core}), one can get the one-point function of $\cO$ in the
 presence of the source $\phi_0$ as
  \be
  \langle \cO(x)\rangle_{\phi_0}=-\frac{\d S_{\mbox{\tiny grav.}}}{\d
  \phi_0(x)}=-\lim_{r\to \infty} \Pi_E(r,x)|_{\phi_E}.
  \ee
 Here $\Pi_E$ is the canonical momentum conjugate to $\phi$ with respect to a foliation in the $r$-direction.
 Namely, $\Pi_E$ is the canonical momentum, taking $r$ as ``time".
 Note also that $\Pi_E$ should be evaluated at the classical
 solution $\phi_E$. Transforming to momentum space, we have
 \be
 \langle \cO(\o_E,\vec{k})\rangle_{\phi_0}=-\lim_{r\to \infty}
 \Pi_E(r,\o_E, \vec{k})|_{\phi_E}.
 \ee
Since $\cO(\o_E,\vec{k})$ is the response of the system to
external perturbations generated by adding the term $\int_{\p
M}\phi_0(x){\cal O}(x)$ to the boundary theory, we have
 \be
 \langle
 \cO(\o_E,\vec{k})\rangle_{\phi_0}=G_E(\o_E,\vec{k})\phi_0(\o_E,\vec{k}).
 \ee
 This allows us to read the Green's function in Euclidean
 signature:
 \be\label{GE}
 G_E(\o_E,\vec{k})=-\large(\lim_{r\to \infty}\frac{\Pi_E(r,\o_E,
 \vec{k})|_{\phi_E}}{\phi_E(r,\o_E,
 \vec{k})}\large)|_{\phi_0=0}.
 \ee
This prescription has no ambiguity since in Euclidean space,
$\phi_E$ is uniquely determined by regularity and the boundary
condition.

In Minkowski spacetime, one can define the Green's function by
doing analytic continuation directly in the zero-temperature case,
as the Euclidean and Minkowski Green's functions are closely
related to each other.  However, for the finite temperature case,
the things become much subtler. In Minkowski space, the retarded
propagator is analytic in the upper half complex-$\o$ plane and is
related to the Euclidean propagator by the relation
 \be\label{EMrelation}
 G_E(\o_E,\vec{k})=-G_R(i\o_E,\vec{k}), \quad \o_E>0.
 \ee
 This means that the value of the retarded propagator along the
 upper imaginary $\o$-axis gives the Euclidean propagator. The
 relation (\ref{EMrelation}) applies for generic cases, not only for  zero temperature
 but also for  finite temperature, and for both bosonic and
 fermionc propagators. For the finite temperature case, $\o_E$ take
 discrete values, and
  \be
 G_E(2\pi Tn,\vec{k})=-G_R(i2\pi Tn,\vec{k}).
 \ee
 On the other hand, for $\o_E<0$, the Euclidean propagator is
 related to the advanced propagator:
  \be
 G_E(\o_E,\vec{k})=-G_A(i\o_E,\vec{k}), \quad \o_E< 0.
 \ee

It seems that one can obtain real-time retarded Green's function
from the inverse relation of (\ref{EMrelation}):
 \be\label{analytic}
 G_R(\o,\vec{k})=-G_E(\o_E,\vec{k})|_{\o_E=-i(\o+i\epsilon)}.
 \ee
For the finite temperature case, this analytic continuation is
quite tricky since $\o_E$ takes only discrete values. One may
ignore this subtlety and just do analytic continuation and wish
the best. This turns out to be the right way. From (\ref{GE}) and
(\ref{analytic}), one may have the retarded correlators from
gravity.

However, it was proposed in \cite{Iqbal:2009fd} that there is an
intrinsic way to obtain the retarded correlators. Instead of
working in Euclidean space, one can work with the classical
solution in Lorentz-signature spacetime. More precisely, one has
the prescription:
 \be\label{GR}
 G_R(\o, \vec{k})=\lim_{r\to \infty}\frac{\Pi(r,\o,
 \vec{k})|_{\phi_R}}{\phi_R(r,\o,
 \vec{k})},
 \ee
 where $\Pi$ is the canonical momentum conjugate to $\phi$, taking $r$ as the ``time"
 direction. Now $\phi_R$ is the classical solution, which should
 be purely in-falling at the black hole horizon and turns to $\phi_0(\vec{k})$
 asymptotically. For the advanced correlators, one has to instead choose
 out-going boundary condition at the horizon.

There is another subtlety in applying the above prescription
(\ref{GR}) to get a finite retarded correlators. In fact, the
naive application of (\ref{GR}) leads to a divergent result. In
order to cancel the divergence,  an extra factor, being powers of
$r$, has to be taken into account. The exact value of the power
depends on what kind of source we consider. For example, if the
asymptotic behavior of the scalar is \be\label{asym}
 \phi\sim A r^{h_R-1}+Br^{-h_R},
 \ee
 where $h_R$ is the conformal weight, and both terms are renormalizable
 if $h_R<1$ such than both $A$ and $B$ can be the source, then the extra factor could
 be $r^{2(h_R-1)}$ or $r^{-2h_R}$, up to which term is taken to be
 the source.

For the scalar perturbation, the conjugate momentum of $\phi$ is
 \be
 \Pi_\phi= -\sqrt{-g}g^{rr}\p_r\phi.
 \ee
For the Maxwell field $A_\mu$, its conjugate momentum is
 \be
 \Pi^\mu=-\sqrt{-g}F^{r\mu}.
 \ee
 For the tensor perturbation, its conjugate momentum is the
 Brown-York stress tensor:
 \be
 \Pi^{\mu\nu}=\frac{\sqrt{-\g}}{16\pi
 G_N}(K^{\mu\nu}-\g^{\mu\nu}K^\l_\l),
 \ee
 where $\g_{\mu\nu}$ and $K_{\mu\nu}$ are the induced metric and
 extrinsic curvature on constant $r$-slice.
For the spinor perturbation, the prescription is quite similar but
the expression of the conjugate momentum depends on the choice of
Gamma matrices.

For the scalar and vector case, the above prescription (\ref{GR})
has turned out to be equivalent to the ones proposed in
\cite{Son05}. And they have been applied to the study the
transport properties of finite temperature ${\cal N}=4$
super-Yang-Mills.

\section{Green's functions in 2D CFT}

In general, it is hard to compare the retarded correlators
obtained from gravity with the ones in CFT.  However, in our case
this is feasible. This is because on one hand the computations in
gravity could be done analytically as we will show, and more
importantly on the other hand the dual two-dimensional CFT is much
more restricted by the symmetries.  In a two-dimensional CFT,
there are two independent sectors: left-moving one and
right-moving one. This requires us to analyze them separately. The
relation between the retarded Green's function and the Euclidean
Matsubara propagator is modified to be
 \be\label{2Drelation}
 G_R(i\o_L,i\o_R)=G_E(\o_{L,E},\o_{R,E}),
 \ee
 at
 \be\label{discrete}
 \o_{L,E}=2\pi n_LT_L,\hs{5ex}\o_{R,E}=2\pi n_RT_R
 \ee
 with $n_L,n_R$ being integers.

In a 2D conformal field theory(CFT), one can define a two-point
function as
 \be
 G(t^+,t^-)=\langle {\cal O}^\dagger_\phi(t^+,t^-){\cal
 O}_\phi(0)\rangle,
 \ee
 where $t^+,t^-$ are the left and right moving coordinates of 2d
 worldsheet, and ${\cal O}_\phi$ is the operator corresponding to
 the field perturbing the black hole.
 For our later use, let us consider an operator of conformal dimensions $(h_L,h_R)$, right charge
 $q_R$, at temperature $(T_L,T_R)$ and chemical potential $\O_R$.
 Its two-point function is decided by conformal invariance\cite{Cardy:1984bb}:
 \be
 G(t^+,t^-)\sim (-1)^{h_L+h_R}\large(\frac{\pi T_L}{\sinh(\pi T_L
 t^+)}\large)^{2h_L}\large(\frac{\pi T_R}{\sinh(\pi T_R
 t^-)}\large)^{2h_R}e^{iq_R\O_R t^-}.
 \ee
For the left mover, it contributes
 \be
 G_E(\o_{L,E})\sim \int_0^{1/T_L}e^{i\o_{L,E}\t_L}\left(\frac{\pi
 T_L}{\sin(\pi T_L\t_L)}\right)^{2h_L}d\t_L,
 \ee
 where $\t_L$ is Euclidean time. In fact, the above function is
 only defined at the discrete frequency in (\ref{discrete}).
 Through analytic continuation, we have
 \be
 G_E(\o_{L,E})\sim
 \frac{T^{2h_L-2}e^{i\o_{L,E}/2T_L}\G(1-2h_L)}{\G(1-h_L+\frac{\o_{L,E}}{2\pi
 T_L})\G(1-h_L-\frac{\o_{L,E}}{2\pi
 T_L})}.\label{left}
 \ee
 For the right mover, it has similar contribution, taken into
 account the extra dependence on the chemical potential:
 \be
 G_E(\o_{R,E})\sim
 \frac{T^{2h_R-2}e^{(i\o_{R,E}+q_R\O_R)/2T_R}\G(1-2h_R)}{\G(1-h_R+\frac{\o_{R,E}-iq_R\O_R}{2\pi
 T_R})\G(1-h_R-\frac{\o_{R,E}-iq_R\O_R}{2\pi
 T_R})}.\label{right}
 \ee
The total contribution is the product of the left-mover's
(\ref{left}) and the right-mover's (\ref{right}):
 \be\label{total}
 G_E(\o_{L,E},\o_{R,E})\sim \frac{T^{2h_L-2}T^{2h_R-2}e^{i\o_{L,E}/2T_L+(i\o_{R,E}+q_R\O_R)/2T_R}\G(1-2h_L)\G(1-2h_R)}
 {\G(1-h_L+\frac{\o_{L,E}}{2\pi
 T_L})\G(1-h_L-\frac{\o_{L,E}}{2\pi
 T_L})\G(1-h_R+\frac{\o_{R,E}-iq_R\O_R}{2\pi
 T_R})\G(1-h_R-\frac{\o_{R,E}-iq_R\O_R}{2\pi
 T_R})}.
\ee

 The CFT absorption cross
 section could be defined with the two-point functions, following
 Fermi's golden rule:
 \be
 \s_{abs}\sim \int dt^+dt^-
 e^{-i\o_Rt^--i\o_Lt^+}[G(t^+-i\epsilon,
  t^--i\epsilon)-G(t^++i\epsilon, t^-+i\epsilon)]
 \ee
Then after being changed into momentum space, the absorption cross
section is
 \be\label{2Dsection}
 \s \sim
 T_L^{2h_L-1}T_R^{2h_R-1}\sinh(\frac{\o_L}{2T_L}+\frac{\o_R-q_R\O_R}{2T_R})
 |\G(h_L+i\frac{\o_L}{2\pi T_L})|^2|\G(h_R+i\frac{\o_R-q_R\O_R}{2\pi
 T_R})|^2.
 \ee

\section{Spacelike warped case}

 The spacelike stretched $AdS_3$ spacetime is the vacuum solution
of three-dimensional topological massive
gravity\cite{Deser:1981wh,Deser:1982vy}. It could be described by
the metric of the form \be\label{spacelikemetric}
 ds^2=\frac{l^2}{v^2+3}[-(1+r^2)d\tau^2+\frac{dr^2}{1+r^2}+\frac{4v^2}{v^2+3}
 (dx+rd\tau)^2],
 \ee
 where $-l^{-2}$ is a negative cosmological constant and the
parameter $v$ is defined to be
 $v\equiv \mu l/3$ with $\mu$ being the mass of the graviton. It
 turns out that only when $v>1$, the spacetime is free of
 pathology\cite{Andy08} and could be a stable vacuum with appropriate
 boundary conditions\cite{Anninos:2009zi}.
This spacetime has $SL(2)_R\times U(1)_L$ isometry group.  Just as
the BTZ black hole\cite{BTZ} could be constructed as discrete
quotient of the $AdS_3$ spacetime, the black hole asymptotic to
spacelike warped $AdS_3$ could be constructed from discrete
identification as well, c.f. \cite{Andy08}.

 The metric of the spacelike stretched warped $AdS_3$ black hole
takes the following form in terms of the Schwarzschild
coordinates:
  \bea
    ds^2=l^2(\textrm{d}t^2+2M(r)\textrm{d}t\textrm{d}\theta+N(r)\textrm{d}\theta^2+D(r)\textrm{d}r^2)
    \label{metric}
 \eea
where \bea
 M(r)&=& v r-\frac{1}{2}\sqrt{r_+r_-(v^2+3)},\\
 N(r)&=&\frac{r}{4}\left(3(v^2-1)r+(v^2+3)(r_++r_-)-4v\sqrt{r_+r_-(v^2+3)}\right),\\
 D(r)&=&\frac{1}{(v^2+3)(r-r_+)(r-r_-)},
 \eea
 Just like the BTZ black hole, there are two horizons located at
 $r=r_+$ and $r=r_-$. We will focus on the physical black holes without any pathology,
 which requires $v > 1$. When $v=1$, there is no stretching and
 the above black hole becomes the usual BTZ black hole, in a rotational frame.

 From discrete identification which leads to the black hole, one can define two temperatures
 \bea\label{tempwarped}
  T_L&=&\frac{(v^2+3)}{8\pi
  l}\left(r_++r_--\frac{\sqrt{(v^2+3)r_+r_-}}{v}\right), \\
  T_R&=&\frac{(v^2+3)(r_+-r_-)}{8\pi
  l},
 \eea
 which are identified to be the temperatures of left-moving and right moving sectors
 in the dual two-dimensional CFT\cite{Andy08}.
 In terms of them, the temperature of the black holes were
 rewritten as
  \be
   \frac{1}{T_H}=\frac{4\pi vl}{v^2+3}\frac{T_L+T_R}{T_R}.
 \ee
 The entropy of the black hole could be recovered from dual CFT
 by applying Cardy formula as well. This help us to determine the
 central charges of dual CFT
  \be
  c_L=\frac{l}{G}\frac{4v}{v^2+3}, \hspace{5ex}
  c_R=\frac{l}{G}\frac{5v^2+3}{v(v^2+3)}.
  \ee
It has been shown in \cite{Compere:2008cv, Blagojevic:2009ek} that
the above central charges could be obtained from central extended
Virasoro algebra, based on the fact that the asymptotic symmetries
of the geometries form a semi-product of a Virasoro algebra and a
current algebra.

The spacelike warped AdS/CFT correspondence states that $v>1$
quantum topological massive gravity with asymptotical spacelike
stretched AdS$_3$ geometry is holographically dual to a
two-dimensional conformal field theory with central charges
$(c_L,c_R)$.

\subsection{Scalar correlators}

 The scalar perturbation about the warped black hole background obeys the
equation of motion: \be
  (\nabla_\mu\nabla^\mu-m^2) \Phi=0.
  \ee
  Since the background (\ref{metric}) has the translational isometry along $t$
  and $\th$, we may make the following ansatz
  \be
  \Phi=e^{-i\o t+ik\th}\phi.
  \ee
With the variable \be z=\frac{r-r_+}{r-r_-}, \label{z}\ee  the
equation of motion on $\phi$ turns out to be \be\label{radial}
 z(1-z)\frac{d^2\phi}{dz^2}+(1-z)\frac{d\phi}{dz}+\frac{1}{(v^2+3)^2}\left(\frac{A}{z}+B+\frac{C}{1-z}\right)\phi=0,
 \ee
where
 \bea
 A&=&\frac{1}{(r_+-r_-)^2}\big(2k+\o\sqrt{r_+}(2v\sqrt{r_+}-\sqrt{v^2+3}\sqrt{r_-})\big)^2,\label{A}
 \\
 B&=&-\frac{1}{(r_+-r_-)^2}\big(2k+\o\sqrt{r_-}(2v\sqrt{r_-}-\sqrt{v^2+3}\sqrt{r_+})\big)^2,\label{B}\\
 C&=&3(v^2-1)\o^2-m^2l^2(v^2+3).
 \eea
The solutions to the equation (\ref{radial}) take the form of
hypergeometric function. Near the horizon, there are two
independent solutions. Only one of them satisfies the purely
ingoing boundary condition at the horizon, which is necessary to
get the retarded Green's function. It is
\be\label{scalarsolution}
 \phi=z^\a (1-z)^\b F(a,b,c,z),  \ee
 where
 \bea
 \a&=&-i\frac{\sqrt{A}}{v^2+3}, \nn\\
 \b&=&\b_\pm=\frac{1}{2}\left(1\pm \sqrt{1-\frac{4C}{(v^2+3)^2}}\right),
 \label{beta}
 \eea
 and
 \bea
 c&=&2\a+1,\nn\\
 a&=&\a+\b+i\sqrt{-B}/(v^2+3),\nn\\
 b&=&\a+\b-i\sqrt{-B}/(v^2+3).\nn
 \eea

For a scalar of mass $m$ propagating in spacelike warped AdS$_3$,
its conformal weight is\cite{{ChenXu2},Anninos2009}
 \be
 h_R^\pm=\D_s^\pm=\frac{1}{2}\pm\sqrt{\frac{1}{4}+s_s},
 \ee
 with
 \be\label{s}
 s_s=\frac{3(1-v^2)}{4v^2}\tilde{k}^2+\frac{l^2}{v^2+3}m^2.
 \ee
Here $\tilde{k}$ is the quantum number with respect to the
translational symmetry along $x$ in the background
(\ref{spacelikemetric}). The presence of quantum number
$\tilde{k}$ in the conformal weight has interesting physical
implications. Obviously, $s_s$ could be negative in a natural way.

   Note that in order to have
a well-behaved asymptotic behavior, we have $\D_s^\pm >0$. More
precisely, if $s_s>0$, we can only choose
 \be
  \D_s^+>1.
 \ee
 On the other side, if $-\frac{1}{4}< s_s\leq 0$, we are free to choose both
 $\D_s^\pm$ with
\be
 0\leq \D_s^-<\frac{1}{2}, \hspace{3ex} \frac{1}{2}< \D_s^+\leq 1.
 \ee
It is remarkable that after taken into account of the subtle
identification of the quantum numbers, $\b$ is directly related to
the conformal weight $\b_\pm=\D_s^\pm$. But note that $\b_\pm$ do
not need to be always positive.

Actually, there is a subtlety in choosing $\b_\pm$ in the above
solution. If $s_s$ is positive, we take the conformal dimension of
the scalar to be $h_R^+>1$, then we have to choose $\b_-$ in
(\ref{scalarsolution}). Correspondingly, the extra factor plugged
in (\ref{GR}) should be $r^{2(h_R^+-1)}$. However, if $s_s<0$ we
have to choose $\b_+$ in the solution. In this case, there are two
normalizable sources, which need special care as we show soon.

 Let us first consider
the source of a dimension $h_R^+>1$. In this case, the momentum is
 \be
 \Pi_s=-\frac{1}{2}(r_+-r_-)(v^2+3)z\p_z \phi
 \ee
 where $\phi=z^\a (1-z)^{\b_-} F(a,b,c;z)$. Note that now
 $\b_-<0$. From the relation
 \bea
 F(a,b,c;z)&=&
 \frac{\G(c)\G(c-a-b)}{\G(c-a)\G(c-b)}F(a,b,a+b-c+1;1-z)\nn\\
 &+&(1-z)^{c-a-b}\frac{\G(c)\G(a+b-c)}{\G(a)\G(b)}F(c-a,c-b,c-a-b+1;1-z),
 \eea
 asymptotically the dominant contribution in $\phi$ is
proportional to $r^{-\b_-}$ since the other term proportional to
$r^{\b_--1}$ runs to zero. On the other hand, in the momentum,
there are various terms asymptotically proportional to
 \be
 r^{-\b_-}, \hs{3ex} r^{\b_--1},\hs{3ex} r^{1-\b_-},\hs{3ex}
 r^{\b_-}.
 \ee
The terms proportional to $r^{\b_-}$ should be picked out. Then
the retarded correlator is just
 \bea
 G_R&\sim& \frac{\G(a+b-c+1)}{\G(c-a-b)}\frac{\G(c-a)\G(c-b)}{\G(a)\G(b)}\nn\\
 &=&{\cal
 N}\frac{\G(2\b_-)}{\G(1-2\b_-)}|\G(c-a)\G(c-b)|^2,\label{retarded}
 \eea
 where
 \be
 {\cal
 N}=\frac{1}{2\pi^2}(\cosh(\frac{2\sqrt{-B}\pi}{v^2+3})-\cos(2\pi\b_-)\cosh(2i\a\pi)+i\sin(2\pi\b_-)\sinh(2i\a\pi)).
 \ee

In order to compare with the CFT result, we have to apply the
identification of quantum numbers to simplify the above relation.
The identification is essential to set up the dictionary of warped
AdS/CFT correspondence. It was suggested in \cite{ChenXu2} that
 \be
 \tilde{k}=\frac{2v}{v^2+3}\o,\hs{3ex}
 \tilde{\o}=\frac{2}{v^2+3}k,\label{identspace}
 \ee
 where $\tilde{k},\tilde{\o}$ are the quantum numbers of global
 warped AdS$_3$ spacetime. $\tilde{\o}$ is the quantum number with respect to the
 translational symmetry along $\tau$ in the background (\ref{spacelikemetric}).
Because that the spacelike warped AdS/CFT correspondence
 is between the spacelike
 stretched  warped $AdS_3$ and its holographically dual 2D CFT,
 we need to use the quantum numbers in global warped AdS$_3$ spacetime rather
 than the ones in the black holes to set up the dictionary. In terms of the quantum numbers
 $\tilde{k},\tilde{\o}$, we have
 \bea\label{betam}
 c-a&=&h_R^+-i\frac{1}{2\pi T_R l}\left(\frac{v^2+3}{2}\tilde{\o}+2\pi
 T_L l\tilde{k}\right), \nn\\
 c-b&=&h_R^+-i\tilde{k},\nn\\
 a&=&1-h_R^+-i\tilde{k},\nn\\
 b&=&1-h_R^+-i\frac{1}{2\pi T_R l}\left(\frac{v^2+3}{2}\tilde{\o}+2\pi
 T_L l\tilde{k}\right).
 \eea

To compare with the CFT result, it is better to rewrite the
correlator in the following form:
 \bea
 G_R\sim {\cal N'}\frac{\G(1-2h_R^+)\G(1-2h_L^+)}{\G(1-h_L^++\frac{i\o_{L}}{2\pi
 T_L})\G(1-h_L^+-\frac{i\o_{L}}{2\pi
 T_L})\G(1-h_R^++\frac{i(\o_{R}-q_R\O_R)}{2\pi
 T_R})\G(1-h_R^+-\frac{i(\o_{R}-q_R\O_R)}{2\pi
 T_R})},\nn\\
 \label{scorrelator}
 \eea
with
 \be
 {\cal
 N'}=\frac{4\pi\sin((2h_R^+-1)\pi)}{\cosh\left(\frac{\o_L}{2\pi
 T_L}-\frac{\o_R-q_R\O_R}{2\pi
 T_R}\right)-\cos\left(2h^+_R-i\left(\frac{\o_L}{2\pi
 T_L}-\frac{\o_R-q_R\O_R}{2\pi
 T_R}\right)\right)}.
 \ee
Here $h^+_L=h^+_R$ and
 \be\label{identsection}
 \o_L=2\pi T_L \tilde{k},\hs{3ex}\o_R=\frac{v^2+3}{2l}\tilde{\o},
 \hs{3ex}q_R=-\tilde{k}, \hs{3ex}\O_R=2\pi T_L.
 \ee
Obviously the retarded correlator (\ref{scorrelator}) is
proportional to (\ref{total}), up to a normalization factor, taken
the relation (\ref{2Drelation}) into account. This fact shows that
real-time scalar correlator in the spacelike stretched AdS$_3$
black hole is well consistent with the prediction of CFT.

The quasi-normal modes of the black hole correspond to the poles
of the retarded Green's function. From (\ref{retarded}), we obtain
that
 \bea
 \o_L&=&-i2\pi T_L(n_L+h_L)\label{spaceleft}\\
 \o_R&=&q_R\O_R-i2\pi T_R(n_R+h_R)
 \eea
 where $n_L$ and $n_R$ are the non-negative integers. This is in precise match with the
 result found in \cite{ChenXu2}.

Moreover,  the imaginary part of real-time retarded correlators
could be identified as the cross section. The cross section reads
 \be\label{section2}
 \s \sim
 \frac{1}{(\G(1-2\b_-))^2}|\G(c-a)\G(c-b)|^2\sinh(2i\a\pi).
 \ee
It could be rewritten as
 \be\label{section3}
 \s \sim \sinh\left((\frac{\o_L}{2\pi T_L}+\frac{\o_R-q_R\O_R}{2\pi T_R})\pi\right)
 |\G(h_L^++i\frac{\o_L}{2\pi T_L})|^2
 |\G(h_R^++i \frac{\o_R-q_R\O_R}{2\pi T_R})|^2,
 \ee
 The relation (\ref{section3}) is reminiscent of the absorption
 cross section in a CFT. It is actually proportional to (\ref{2Dsection}),
 up to a normalization factor. The
situation is very similar to the one in Kerr/CFT
correspondence\cite{Bredberg:2009pv}. The normalization factor
depends on the temperatures as
 \be
 (T_L)^{h_L}(T_R)^{h_R}.\nn
 \ee
Such a factor may arise from the nontrivial coordinate
transformation, which redefine the temperature, as in BTZ
case\cite{Iqbal:2009fd}.

There is another interesting issue on the cross section of a
scalar scattering the black hole. It has been argued that for a
massless scalar, the low energy limit of the cross section is
proportional to the horizon area\cite{Das:1996we}. For the
spacelike stretched black hole, this has been studied in
\cite{Oh:2008tc,Kao:2009fh}. In our study, since we cannot fix the
normalization factor exactly, we can not check this statement
precisely. However, we can check if the low energy limit of the
cross section has the right dependence on frequency. It is
straightforward to investigate the $\omega \to 0$ limit, with
$m=k=0$, which reads $\s \sim \o$. This linear dependence on $\o$
is in consistent with the universal statement.

It is more tricky if $s_s<0$ so the scalar field can pick both the
conformal weight $h_R^\pm$. In this case, note that one has to
choose $\b_+$ in the scalar solution.  The asymptotic behavior of
the scalar field is like
 \be
 \phi\sim Ar^{-\b_+}+Br^{-1+\b_+}.
 \ee
Since $0<\b_+<1$, both terms are normalizable. We are free to
choose either one as the source: $A$ is the source of a dimension
$1-\b_+=h_R^-$, while $B$ is the source of a dimension
$\b_+=h_R^+$. In other words, if we want to study the correlator
of the operators of dimension $h_R^-$, then we should find the
term proportional to $r^{\b_+}$ in the canonical momentum $\pi_s$.
Consequently, the retarded correlator take the same form as
(\ref{retarded}), with $\b_-$ being replaced by $\b_+$. As a
result of the replacement, the conformal dimension appeared in the
retarded correlator (\ref{scorrelator}) and the cross section
(\ref{section3}) should be $h_R^-$ rather than $h_R^+$.

On the other hand, in order to study the correlator of the
operator of dimension $h_R^+$,  we have to find the term
proportional to $r^{1-\b_+}$ in the canonical momentum $\Pi_s$.
Consequently, the retarded Green's function is now
 \bea\label{retarded2}
 G_R &\sim&
 \frac{\G(a)\G(b)}{\G(c-a)\G(c-b)}\frac{\G(c-a-b)}{\G(a+b-c)}\nn\\
 &=&{\cal N_1}\frac{\G(1-2\b_-)}{\G(2\b_--1)}|\G(a)\G(b)|^2,
 \eea
 where
 \be
 {\cal
 N_1}=\frac{1}{2\pi^2}(\cosh(\frac{2\sqrt{-B}\pi}{v^2+3})-\cos(2\pi\b_-)\cosh(2i\a\pi)
 +i\sin(2\pi\b_-)\sinh(2i\a\pi)).
 \ee

Taking into account of the identifications of quantum numbers
(\ref{identspace}), we find that
 \bea\label{betap}
 b&=&h_R^+-i\frac{1}{2\pi T_R l}(\frac{v^2+3}{2}\tilde{\o}+2\pi
 T_L l\tilde{k}), \nn\\
 a&=&h_R^+-i\tilde{k},\nn \\
 c-b&=&1-h_R^+-i\tilde{k},\nn \\
 c-a&=&1-h_R^+-i\frac{1}{2\pi T_R l}(\frac{v^2+3}{2}\tilde{\o}+2\pi
 T_L l\tilde{k}).
 \eea
Note that it is different from (\ref{betam}). The difference stems
from the choice of $\b_\pm$ in the scalar solutions. Nevertheless,
the retarded correlator takes exactly the same form as
(\ref{scorrelator}), even though the scalar solution now is
different. In this case, the poles of the real-time correlator is
simply
 \be
 a=-n_L, \hs{5ex} \mbox{or} \hs{2ex} b=-n_R,
 \ee
 with $n_L,n_R$ being non-negative integer. This is the same as the
 ones found in \cite{ChenXu2}.

Similarly the cross section is
 \bea
 \s&=&Im(G_R)\sim
   \frac{1}{\G(2\b_+)\G(2\b_+-1)}|\G(a)\G(b)|^2\sinh(2i\a\pi).
 \eea
It could be written as (\ref{section3}).  And the comparison with
the prediction of CFT is along the similar line. The low energy
cross section for massless scalar could be obtained by taking
$m\to 0,
 \o \to 0$ limit and is found to be proportional to $\o$ as well.

In short, no matter what kind of scalar source we choose, we end
up with the same real-time retarded correlator (\ref{scorrelator})
and the cross section (\ref{section3}). These results are in good
match with the prediction (\ref{total},\ref{2Dsection}), up to a
normalization factor.

Before ending this section, we would like to discuss another
interesting case when the black hole has super-radiance. This
happens when
 \be
 S=\frac{1}{4}+\frac{3(1-v^2)}{4v^2}\tilde{k}^2+\frac{l^2}{v^2+3}m^2<
 0.
 \ee
 This is possible now due to the presence of the quantum number
 $\tilde{k}$ in the conformal weight. Therefore even
 though the mass-square of the scalar field satisfies the
 Breitenlohner-Freedman bound for three-dimensional AdS spacetime,
 the perturbation could still be unstable. As a result,
 superradiance may happen in the spacelike stretched AdS$_3$
 black holes\cite{Anninos2009}, just like the superradiance in Kerr black hole\cite{Bredberg:2009pv}.
 Superradiance happens in the null warped AdS$_3$
 spacetime as well.

 In this case, the conformal weight is just
 \be
 h_R^\pm=\frac{1}{2}\pm i \sqrt{-S}
 \ee
 No matter which conformal weight and the corresponding eigenfunction we choose, the
 asymptotic behavior of the scalar field is always of the form
 \be
 \phi \sim S_1 (1-z)^{\frac{1}{2}+i\sqrt{-S}} + S_2
 (1-z)^{\frac{1}{2}-i\sqrt{-S}}.
 \ee
 The first term could be taken as the ingoing wave and the second term
as the outgoing wave. The ratio $S_2/S_1$ measures the response of
the black hole to an incoming wave, and is related to the retarded
two-point function of dual CFT. Therefore, we have
 \be\label{retardsuper}
 G_R \sim \frac{S_2}{S_1}=\left\{\begin{array}{l}
 \frac{\G(c-a)\G(c-b)\G(a+b-c)}{\G(a)\G(b)\G(c-a-b)},\hs{3ex}\mbox{if
 $h_R=h_R^+$}\\
 \frac{\G(a)\G(b)\G(c-a-b)}{\G(c-a)\G(c-b)\G(a+b-c)},\hs{3ex}\mbox{if $h_R=h_R^-$.}
 \end{array}\right.\ee
 Note that since even with a complex conformal weight the relations
 (\ref{left},\ref{right}) always make sense,  the retarded
 correlator (\ref{retardsuper}) is still consistent with (\ref{2Drelation},\ref{total}).

\subsection{Vector correlator}

The equation of motion for the massive vector field in three
dimension could be cast into a first-order differential equation
of the form
 \be\label{vectoreq}
   \epsilon_{\lambda}^{\
   \alpha\beta}\partial_{\alpha}A_{\beta}=-mA_{\lambda},\label{vector}
  \ee where $\epsilon_{\lambda}^{\ \alpha\beta}$ is the Levi-Civita
tensor with $\epsilon^{tr\theta}=1/\sqrt{-g}$. The Killing
symmetry of the background allows us to make the following ansatz:
\be
    A_{\mu}=e^{-i\omega t+ik\theta}\phi_{\mu}.
\ee The complete solution could be found in \cite{ChenXu2}:
 \bea
 \phi_t&=&z^{\alpha_v} (1-z)^{\beta_v+1} F(a_v+1,b_v+1,c_v,z),
 \nn\\
 \phi_\theta&=&\tilde A_v\phi_t+\tilde B_v
\frac{1}{1-z}\phi_t+\tilde C_v
z\frac{\textrm{d}\phi_t}{\textrm{d}z},\nn\\
 \phi_r&=&-\frac{2D(r)}{m l}(ik\phi_t+i\omega\phi_\theta),
  \eea
where $\alpha_v=-i\sqrt{A_v},\ \beta_v=(-1+\sqrt{1-4C_v})/2$ and
 \be c_v=1+2\alpha_v,\ \ a_v=\alpha_v+\beta_v+i\sqrt{-B_v},\ \
 b_v=\alpha_v+\beta_v-i\sqrt{-B_v},
 \ee
 with
 \bea
A_v&=&\frac{1}{(r_+-r_-)^2(v^2+3)^2}\left(2k+\omega\sqrt{r_+}(2v\sqrt{r_+}-\sqrt{v^2+3}\sqrt{r_-})\right)^2,\nn
 \\
 B_v&=&-\frac{1}{(r_+-r_-)^2(v^2+3)^2}\left(2k+\omega\sqrt{r_-}(2v\sqrt{r_-}-\sqrt{v^2+3}\sqrt{r_+})\right)^2,\nn\\
 C_v&=&\frac{1}{(v^2+3)^2}\left(3(v^2-1)\omega^2-(m^2l^2+2mv
 l)(v^2+3)\right),\\
\tilde A_v&=&\frac{1}{2\omega^2+2m^2l^2}(-2\omega k+2m^2l^2vr_--m^2l^2\sqrt{r_+r_-(v^2+3)}),\nn\\
 \tilde B_v&=&\frac{m^2l^2v(r_+-r_-)}{\omega^2+m^2l^2},\nn\\
 \tilde C_v&=&-\frac{m l(v^2+3)(r_+-r_-)}{2\omega^2+2m^2l^2}.\nn
\eea

 The conjugate momentum is the conserved current $\mathcal
  {J}^{\mu}$
  \bea
         \mathcal {J}^{\mu}=-\lim_{r\rightarrow \infty
         }\sqrt{-g}F^{r\mu}.
  \eea
    Using the equation of motion, we find that the current is simply related
    to  the fields
   \be
      \mathcal{J}^{t}=mA_{\th},\ \ \mathcal{J}^{\th}=-mA_{t}
   \ee

   In order to use the prescription (\ref{GR}) for the retarded
   Green's function, we have to analyze the asymptotic behavior of the
   vector fields. It can be read directly from the solutions
   as the following
   \bea
     \phi_t&=&A_1(1-z)^{1+\beta_v}+A_2(1-z)^{-\b_v},\\
     \phi_{\th}&=&(\tilde B_v-\tilde C_v(\b_v+1))A_1(1-z)^{\b_v}
     +(\tilde B_v+\tilde C_v\b_v)A_2(1-z)^{-\b_v-1}
   \eea
 where
   \bea
     A_1=\frac{\G(c_v)\G(c_v-a_v-b_v-2)}{\G(c_v-a_v-1)\G(c_v-b_v-1)}A_0,\
     \ \
     A_2=\frac{\G(c_v)\G(a_v+b_v+2-c_v)}{\G(a_v+1)\G(b_v+1)}A_0,
   \eea
   with $A_0$ being a constant.
   If the dominant term in $\phi_t$ is the source, then the retarded Green's function
  describing the response is given by
   \bea
     G_{tt}=\frac{m\left(\tilde B_v-\tilde
     C_v(\b_v+1)\right)A_1}{A_2}.
   \eea
   If the $\phi_{\th}$ is the source, the corresponding retarded
   Green's function is
   \bea
     G_{\th\th}=-\frac{mA_1}{(\tilde B_v+\tilde C_v\b_v)A_2}.
   \eea

Up to a normalization factor, we have in both cases
 \bea\label{vcorrelator}
 G_R \sim \frac{A_1}{A_2}&=&\frac{\G(a_v+1)\G(b_v+1)\G(c_v-a_v-b_v-2)}
 {\G(c_v-a_v-1)\G(c_v-b_v-1)\G(a_v+b_v+2-c_v)}\nn\\
  &=&{\cal
  N}_v\frac{\G(1-2h_R^v)}{\G(2h_R^v-1)}|\G(a_v+1)\G(b_v+1)|^2,
 \eea
 with
 \be
 {\cal N}_v=\frac{1}{2\pi^2}(\cosh(2\sqrt{-B_v}\pi)-\cos(2\pi h_R^v)\cosh(2i\a_v\pi)
 +i\sin(2\pi h_R^v)\sinh(2i\a_v\pi)).
 \ee
 Here $h_R^v$ is the conformal weight of a massive vector
 field\cite{ChenXu2}
 \be\label{vweight}
 h_R^v=\frac{1}{2}+\sqrt{\frac{1}{4}+s_v}
 \ee
 with
 \be
 s_v=\frac{3(1-v^2)}{4v^2}\tilde{k}^2+\frac{(m^2l^2+2vml)}{v^2+3}.
 \ee
Taking into account of the fact
 \bea
 c_v-a_v-1&=&1-h^v_R-i\frac{1}{2\pi T_R l}(\frac{v^2+3}{2}\tilde{\o}+2\pi
 T_L l\tilde{k}),\nn\\
 c_v-b_v-1&=&1-h^v_R-i\tilde{k},\nn\\
 a_v+1&=&h^v_R-i\tilde{k},\nn\\
 b_v+1&=&h^v_R-i\frac{1}{2\pi T_R l}(\frac{v^2+3}{2}\tilde{\o}+2\pi
 T_L l\tilde{k}),
 \eea
 we find that just like the scalar case, the vector correlator is well
 consistent with (\ref{total}), up to a normalization factor.

 However, strictly speaking there is a discrepancy from the CFT
prediction. Now for the vector field, the conformal weights of
left-mover and right-mover are different:
 \be
 h_L^v=h_R^v\pm 1.
 \ee
 Correspondingly the contribution from left-mover is slightly different
 from the right-mover one. However, this difference could not be
 seen in (\ref{vcorrelator}), recalling the fact that both
 the real parts of $(1+a_v)$ and $(1+b_v)$ give the same $h_R^v$.
 Nevertheless,  using $\G(1+z)=z\G(z)$, one can absorb the extra
 $a_v$ factor into the normalization factor such that one has
 \be
 G_R \sim {\cal N}_v^\prime\frac{1}{|\G(c_v-a_v)\G(c_v-b_v-1)|^2}.
 \ee
 Then the parts involving the Gamma functions have the right dependence
 consistent with the prediction of CFT. It is remarkable that
 the normalization factor including the contribution of $a_v$ is
 independent of $\tilde{\o}$ so that the retarded correlator has
 the correct frequency dependence for the Green's function of a
 finite temperature CFT.

The cross section can be read directly
 \be\label{sectionvec}
 \s \sim
 \frac{1}{\G(2h_R^v)\G(2h_R^v-1)}|\G(a_v+1)\G(b_v+1)|^2\sinh(2i\a_v\pi)).
 \ee
It is similar to (\ref{section3}) in the scalar case, with $h_R^v$
replacing $h_R^+$. The discussions on the quasi-normal modes, the
low energy limit of massless vector cross section and
super-radiance are very similar to the scalar case. We just omit
them here.

\section{Null warped case}

The null warped $AdS_3$ spacetime is another vacuum solution of
three-dimensional topological massive gravity. It is only well
defined at $v=1$. Similar to other warped $AdS_3$ spacetimes, it
also has isometry group $SL(2,R)\times U(1)_{null}$. The null
warped black holes could be obtained as the quotient of the null
warped $AdS_3$. The  metric of the null warped black hole is of
the form
 \be\label{nullmetric}
   \frac{ds^2}{l^2}=-2r\textrm{d}\theta\textrm{d} t+(r^2+r+\alpha^2)\textrm{d}
   \theta^2+\frac{\textrm{d} r^2}{4r^2}
 \ee
 where $1/2>\alpha>0$ in order to avoid the naked causal
 singularity. The horizon is located at $r=0$.
 From the thermodynamics of this extremal black hole, it was argued that
 there exist only non-vanishing right-moving temperature
 \be
 T_R=\frac{\a}{\pi l}.
 \ee

There is a conjecture on the null warped background\cite{ChenXu2}:
$v=1$ quantum topological massive gravity with asymptotical null
warped $AdS_3$ geometry is holographically dual to a 2D boundary
CFT with the left-moving central charge
$c_L=\frac{l}{G}\frac{4v}{v^2+3}$ and the right-moving central
charge $c_R=\frac{(5v^2+3)l}{Gv(v^2+3)}$. From the black hole
entropy, it seems that it is not necessary to have left-moving
central charge since $T_L=0$. However, the diffeomorphsim anomaly
requiring that $c_L-c_R=-\frac{l}{Gv}$ asks for the existence of
the left-moving sector.

\subsection{Scalar correlator}

For a scalar field in the null warped black holes background
(\ref{nullmetric}), its solution can be expressed by Kummer
function as
   \be
   \phi_\pm=e^{-i\o t+ik\th}e^{-\frac{z}{2}}z^{\frac{1}{2}\pm\tilde m_s}F(\frac{1}{2}\pm\tilde m_s-\k,1\pm2\tilde m_s, z)
 \ee
 where $z=-i\o\alpha\frac{1}{r},\ \tilde
 m_s=\frac{1}{2}\sqrt{1+m^2l^2-\o^2}$ and $
 \k=\frac{i}{4\alpha}(\o-2k)$. The wave function should be a
 composition
 \be
 \phi=\tilde{A}_1\phi_+ +\tilde{A}_2 \phi_-.
 \ee
In order to obtain the retarded Green's function, the solutions
should satisfy the purely ingoing boundary condition at the
horizon. This requirement helps us to fix the normalization
factors relatively
 \be
 \tilde A_1=-\frac{\G(1-2\tilde m_s)}{\G(\frac{1}{2}-\tilde
   m_s-\k)}(-i\o\a)^{\frac{1}{2}+\tilde m_s}\tilde A_0,\hspace{4ex}
 \tilde  A_2=\frac{\G(1+2\tilde m_s)}{\G(\frac{1}{2}+\tilde
   m_s-\k)}(-i\o\a)^{\frac{1}{2}-\tilde m_s}\tilde A_0,
\ee where $\tilde A_0$ is a constant, and
$-i=e^{-\frac{i\pi}{2}}$.

The conformal weight of the scalar field of mass $m$ in the null
warped AdS$_3$ spacetime is
 \be
 h^\pm=\frac{1}{2}(1\pm\sqrt{1+s_n})
 \ee
with
\be
 s_n=m^2l^2-4\tilde{k}^2,
\ee where $\tilde{k}$ is the $U(1)$ quantum number of the global
null warped background. As usual, the conformal weight should be
positive for stable perturbation. Similar to the spacelike warped
case, if $s_n>0$, there is only one possible conformal weight
$h^+>1$, while if $-1<s_n<0$ there are two possible choices
$h^\pm$ with $0<h^-<\frac{1}{2}, \frac{1}{2}<h^+<1$.

   The asymptotic behavior of the solution is
\be
   \phi \sim \tilde A_1z^{h}+\tilde A_2z^{1-h}
\ee
   where $s_n>0$ and $h>1$.     The conjugate momentum is
\be
   \Pi_{null}\propto \p_z\phi
 \ee
 In this case, the source term is $\tilde A_2$ and correspondingly in the
 conjugate momentum the relevant terms should be proportional to
 $z^{h-1}$. Then the retarded Green's function is
\bea
   G_R & \sim & \frac{\tilde A_1}{\tilde A_2}\propto -\frac{\G(1-2\tilde m_s)}{\G(1+2\tilde m_s)}\frac{\G(\frac{1}{2}+\tilde
    m_s-\k)}{\G(\frac{1}{2}-\tilde m_s-\k)},\label{retardnull}
\eea

If $s_n<0$, just as usual AdS/CFT case, there could be two kinds
of renormalizable sources. We are free to choose one of them, then
we have
 \be
 G_R\sim \left\{\begin{array}{l}
 \frac{\tilde A_1}{\tilde A_2}, \hs{3ex}\mbox{taking $\tilde A_2$ as the source}\\
 \frac{\tilde A_2}{\tilde A_1}, \hs{3ex}\mbox{taking $\tilde A_1$ as the source.}
 \end{array}\right.
 \ee
In the following discussion, we will focus on the case with
conformal weight $h=\frac{1}{2}+\tilde{m}_s>1$ without losing
generality.

As the spacelike case, we need to take into account of the
identification of quantum numbers, which was discussed in
\cite{ChenXu2}
 \bea\label{identn}
 k&=&-\tilde{\o}, \\
 \o&=&2\tilde{k}.
 \eea
Then we have
 \bea
 \frac{1}{2}+\tilde{m}_s-\k &=&h-\frac{i}{2\pi T_R
 l}(\tilde{k}+\tilde{\o}), \\
 \frac{1}{2}-\tilde{m}_s-\k &=&1-h-\frac{i}{2\pi T_R
 l}(\tilde{k}+\tilde{\o}).
 \eea
Now it is easy to see that the correlator is in good match with
(\ref{right}), once the frequency, the charge and the chemical
potential in the right-moving sector are identified to be
 \be
 \o_R=\frac{\tilde{\o}}{l},\hs{3ex}q_R=-\tilde{k},\hs{3ex}\O_R=\frac{1}{l}.
 \ee
Note that in the null warped case, the absence of the left
temperature make the theory to be ``chiral", so it matches with
the right-moving sector of dual CFT.

The quasi-normal modes correspond to the poles of the retarded
correlator, which is just
 \be
 \frac{1}{2}+\tilde{m_s}-\k=-n.
  \ee
Taking into account of (\ref{identn}), we have
 \be
 \tilde{\o}_R=-\tilde{k}_n-i2\pi T_Rl(n+h_R).
 \ee
 This is exactly we have found in \cite{ChenXu2}.

  The cross section is now
\bea
  \s\sim ImG_R\sim \frac{1}{(\G(2h_R))^2}
  \frac{\pi(h_R-\frac{1}{2})}{\sin(h_R\pi)}\sinh\left(\frac{\tilde{k}+\tilde{\o}}{2T_Rl}\pi\right)
  |\G(h+\frac{i(\tilde{k}+\tilde{\o})}{2\pi T_R l})|^2.
\eea It looks quite similar to the prediction (\ref{2Dsection}) of
dual 2D CFT, up to a prefactor being the power of  the
temperature. In the low energy limit for the massless scalar
  field, we have $ImG_R\sim-l\o\a$. So the cross section $\s$ is
  proportional to $\o \mathcal {A}_H$, where $\mathcal {A}_H=2\pi l
  \a$ is the area of the horizon.

\subsection{Vector correlation}

After making the ansatz $A_\mu=e^{-i\omega
t+ik\theta}\phi_\mu(r)$, we also have two independent solutions of
Eq.(\ref{vectoreq}) as
 \bea
 \phi_{t\pm}&=&e^{-\frac{z}{2}}z^{\frac{1}{2}\pm\tilde m_v}F(\frac{1}{2}\pm\tilde m_v-\k,1\pm2\tilde m_v,
   z)
  \eea
  and $\phi_\th,\ \phi_r$ can be read from
  \bea
 \phi_\th&=& \frac{2mlr^2}{\o^2}(\frac{\textrm{d}\phi_t}{\textrm{d}r}+(\frac{\omega
   k}{2mlr^2}+\frac{ml}{2r})\phi_t),\nn\\
 \phi_r&=&\frac{-1}{2mlr^2}(ik\phi_ t+i\omega\phi_\theta),
 \eea
where $z=-i\omega\alpha \frac{1}{r},\ \tilde
 m_v=\pm\frac{1}{2}\sqrt{(ml-1)^2-\omega^2}$ and $
 \k=\frac{i}{4\alpha}(\omega-2k)$.

   Similarly we have to make a combination of the two solutions to
   satisfy the boundary condition that there are only purely ingoing modes at the
   horizon. We get
    \be
      \phi_\mu=C_1\phi_{\mu+}+C_2\phi_{\mu-}
   \ee
    where
     \bea
       C_1=-\frac{\G(1-2\tilde m_v)}{\G(\frac{1}{2}-\tilde
   m_v-\k)}C, \hspace{4ex}
   C_2=\frac{\G(1+2\tilde m_v)}{\G(\frac{1}{2}+\tilde
   m_v-\k)}C
     \eea
     with $C$ being a constant.
    The  asymptotic solution for the vector field in the null black
   hole background is
   \bea
      \phi_t&=&C_1z^{\frac{1}{2}+\tilde m_v}+C_2z^{\frac{1}{2}-\tilde
      m_v},\\
      \phi_{\th}&=&\frac{i m l\a}{\o}(-1-2\tilde m_v+m l)C_1z^{-\frac{1}{2}+\tilde m_v}
      +\frac{i m l\a}{\o}(-1+2\tilde m_v+m l)C_2z^{-\frac{1}{2}-\tilde
      m_v}.
   \eea

     If we choose $\phi_t$ as the source, then we have the retarded Green
     function
     \bea
       G_{tt}&=&\frac{i m^2 l\a}{\o C_2}(-1-2\tilde m_v+m l)C_1.
            \eea
     Similarly taking the $\phi_\th$ as the source, we have
     \bea
       G_{\th\th}&=&\frac{i\o C_1}{\a l(-1+2\tilde m_v+m l)C_2}.
              \eea
In any case, the retarded Green's function is proportional to
$C_1/C_2$. Similarly, from
 \be
 \frac{1}{2}-\tilde m_v-\k=1-h_R^v-\frac{i}{2\pi T_R
 l}(\tilde{k}+\tilde{\o})
 \ee
 with $h_R^v$ being the conformal weight of massive vector fields
 in the null warped AdS$_3$
 \be
 h_R^v=\frac{1}{2}+ \frac{1}{2}\sqrt{(ml-1)^2-4\tilde{k}_n^2},
 \ee
 we see that the retarded Green's function is consistent with
 (\ref{right}). The quasi-normal modes could be read
directly from the poles:
 \be
 \frac{1}{2}+\tilde
   m_v-\k=-n,
 \ee
 which gives
\be
 \tilde{\o}_R^v=-\tilde{k}_n-i2\pi T_Rl(n+h_R^v).
 \ee
 Similarly, one can read out the
cross section from the retarded Green's function, which is
consistent with the prediction of dual CFT.

\section{Extremal stretched space-like warped black hole}

   In this section, we consider the retarded Green's function for
the extremal stretched space-like warped black hole. The metric of
the black hole is still of the form (\ref{metric}), but with
$r_+=r_-=r_0$. In this case, the right moving temperature is just
zero. However, the eigenfunctions of the perturbations can not be
obtained simply from the non-extremal ones by taking the limit
$r_+=r_-$ since the variable $z$ defined in (\ref{z}) becomes a
constant and $A, B$ given in (\ref{A})(\ref{B}) are divergent. One
has to solve the equations of motion anew. In fact, we will see
soon that the eigenfunctions are  given by Kummer functions rather
than hypergeometric functions. It is very much like the null black
holes which are also extremal black holes.

Certainly, the warped AdS/CFT correspondence still make sense in
the extremal limit. The only thing one has to take care is that
there is only non-vanishing left-moving temperature, but there are
still central charges in both left-moving and right-moving
sectors. This is not a surprise since the asymptotical geometry,
the global warped AdS$_3$ is intact and the black hole changes
only the local geometry.

\subsection{Scalar correlator}

  Let us start from the scalar field
case. Making the  ansatz $\Phi=e^{-i\o t+ik\th}\phi$ and
introducing the
  variable $
  z=\frac{\hat a}{r-r_0}$ with
  \be
  \hat a\equiv -\frac{2i(2\o
  vr_0-\o\sqrt{v^2+3}r_0+2k)}{(v^2+3)},\ee
  we find the equation of motion
  \be
    \frac{d^2\phi}{dz^2}+(\frac{\frac{1}{4}-\hat
    m_s^2}{z^2}+\frac{\hat
    \k}{z}-\frac{1}{4})\phi=0,
  \ee
  where $\hat
  m_s=\sqrt{\frac{1}{4}+\frac{m^2l^2}{v^2+3}-\frac{3(v^2-1)\o^2}{(v^2+3)^2}}$
  and $\hat \k=\frac{2i\o v}{v^2+3}$.
   It has two independent solutions
   \be
     \phi_\pm=e^{-\frac{z}{2}}z^{\frac{1}{2}
     \pm\hat m_s}F(\frac{1}{2}\pm\hat m_s-\hat\k,1\pm2\hat m_s, z)
   \ee
   Obviously the situation is quite similar to the scalar field in the null black
   hole background. The retarded Green's function can be found in the
   similar way. First, we should combine the two solution into a
   solution with purely ingoing mode at the black hole horizon. We
   find
   \be
      \phi=\hat A_1\phi_++\hat A_2\phi_-
   \ee
   where
     \be
    \hat A_1=-\frac{\G(1-2\hat m_s)}{\G(\frac{1}{2}-\hat
   m_s-\hat\k)}\hat A,\hspace{4ex}
   \hat A_2=\frac{\G(1+2\hat m_s)}{\G(\frac{1}{2}+\hat
   m_s-\hat\k)}\hat A,
     \ee
     with $\hat A$ being a constant.
   The conjugate momentum is
  \be
    \Pi=\frac{1}{2}l(v^2+3)\hat a\p_z\phi
  \ee

  Taking into the identification (\ref{identspace}), we see that
  the conformal weights of the scalar field of mass $m$ are
   \be
   h^\pm_R=\frac{1}{2}\pm \hat m_s.
   \ee
   Certainly, it is exactly the same as the ones in the
   non-extremal case.
If we choose $s_s>0$ and $h_R^+>1$, then we get real-time retarded
correlator
 \be
 G_R \propto \frac{\hat A_1}{\hat A_2}.
 \ee
If we have $s_s<0$, then we can choose either $\hat A_1$ or $\hat
A_2$ as source, and we find that
 \be
 G_R\propto \left\{\begin{array}{l}
 \frac{\hat A_1}{\hat A_2}, \hs{3ex}\mbox{taking $\hat A_2$ as the source}\\
 \frac{\hat A_2}{\hat A_1}, \hs{3ex}\mbox{taking $\hat A_1$ as the source.}
 \end{array}\right.
 \ee
 In either case, the Green's function is consistent with the
 prediction (\ref{left}).

Let us focus on the case with the conformal weight $h^+>1$, then
  \be
  G_R \propto
  \frac{A_1}{A_2}=\frac{\G(2-2h)}{\G(2h)}\frac{\G(h^+-i\tilde{k})}{\G(1-h^+-i\tilde{k})}.
 \ee
Note that since $\tilde{k}=\frac{\o_L}{2\pi T_L}$, the correlator
is actually well consistent with (\ref{left}).

 It is interesting to read out the quasi-normal
modes. In either case, they are characterized by the relation
 \be\label{quasiext}
 \tilde{k}=-i(n+h_L),
 \ee
 with $n$ being a non-negative integer and $h_L=h_R$. This
 relation is exactly the same as the one (\ref{spaceleft})
 in the left-moving sector of the
 non-extremal spacelike stretched black hole. As we expected,
 there is no quasi-normal modes in the right-moving sector.

 The cross section is
  \be
      \sigma=Im G_R\sim \frac{\G(2-2h
       )|\G(h-i\tilde\k)|^2}{\G(2h)}\cos(h\pi)\sinh(\tilde{k}\pi).
  \ee
  It is proportional to $\frac{\o\mathcal {A}_H}{2\pi}$ in the low energy limit for the $s$ wave, where $\mathcal{A}_H=\pi
  l(2v-\sqrt{v^2+3})$ is the area of the horizon.

\subsection{Vector correlator}

    Now let us move to the correlators of the vector fields.  Taken the
ansatz \be
    A_{\mu}=e^{-i\omega t+ik\theta}\phi_{\mu},
\ee then the equations of motion can be given explicitly
 \bea
 \frac{\textrm{d}\phi_t}{\textrm{d}r}&=&2D(r)\left((\frac{-\omega k}{m l}+m l M(r))\phi_t-(\frac{\omega^2}{m
 l}+m l)\phi_\theta\right),\label{eom1}\\
 \frac{\textrm{d}\phi_\theta}{\textrm{d}r}&=&2D(r)\left ((\frac{k^2}{m
 l}+m l N(r))\phi_t-(\frac{-\omega k}{m
 l}+m l M(r))\phi_\theta\right),\\
 \phi_r&=&-\frac{2D(r)}{m l}(ik\phi_t+i\omega\phi_\theta)\label{eom2}.
  \eea
     The two independent solutions of $\phi_t$ are
     \be
     \phi_{t\pm}=e^{-\frac{z}{2}}z^{\frac{1}{2}
     \pm\hat m_v}F(\frac{1}{2}\pm\hat m_v-\hat\k,1\pm2\hat m_v,
     z),
     \ee
     where $\hat
  m_v=\sqrt{\frac{1}{4}+\frac{m^2l^2+2mlv}{v^2+3}-\frac{3(v^2-1)\o^2}{(v^2+3)^2}}$
  and $\hat \k,\ z$ are the same as the ones in the scalar field case.
   Now we should also combine these two solutions into a
   solution with purely ingoing boundary condition at the black hole horizon
   \be
      \phi=\hat C_1\phi_++\hat C_2\phi_-
   \ee
   where
     \be
    \hat C_1=-\frac{\G(1-2\hat m_v)}{\G(\frac{1}{2}-\hat
   m_v-\hat\k)}\hat C_0,\hspace{4ex}
   \hat C_2=\frac{\G(1+2\hat m_v)}{\G(\frac{1}{2}+\hat
   m_v-\hat\k)}\hat C_0,
     \ee
with $\hat C_0$ being a constant.

   $\phi_\th,\ \phi_r$ can be obtained directly from the equations
   (\ref{eom1})(\ref{eom2}).  If $\o^2+m^2l^2\neq 0$,
   \be
     \phi_\th=\hat A\phi_t+\frac{\hat B}{z}\phi_t+\hat
     C\p_z\phi_t,
   \ee
   where
   \bea
     \hat A=\frac{m^2l^2(2v-\sqrt{v^2+3})r_0-2wk}{2\o^2+2m^2l^2},\\
     \hat B=\frac{m^2l^2v\hat a}{2\o^2+2m^2l^2},\ \ \
     \hat C=\frac{ml(v^2+3)\hat a}{2\o^2+2m^2l^2}.
   \eea
   The asymptotic behavior of the fields are
   \bea
      \phi_t&=&\hat C_1z^{\frac{1}{2}+\hat m_v}+\hat C_2z^{\frac{1}{2}-\hat m_v},\\
      \phi_\th&=&(\hat B+\hat C(\frac{1}{2}+\hat m_v))\hat C_1z^{-\frac{1}{2}+\hat m_v}
        +(\hat B+\hat C(\frac{1}{2}-\hat m_v))\hat C_2z^{-\frac{1}{2}-\hat
        m_v}.
   \eea
 So if the dominant term in $\phi_t$ is the source, then the retarded Green's function
  describing the response is given by
   \bea
     G_{tt}=\frac{m\left(\hat B-\hat C(\frac{1}{2}+\hat m_v)\right)\hat C_1}{\hat C_2}
   \eea
   If the $\phi_{\th}$ is the source, the corresponding retarded
   Green's function is
   \bea
     G_{\th\th}=-\frac{m\hat C_1}{(\hat B+\hat C(\frac{1}{2}-\hat m_v))\hat C_2}
   \eea

Up to a normalization factor, we have in both cases
 \bea
 G_R &\sim& \frac{\hat C_1}{\hat C_2}={\cal
  N}_v'\frac{\G(1-2\hat m_v)}{\pi\G(1+2\hat m_v)}|\G(\frac{1}{2}+\hat m_v-\hat \k)|^2,\label{retardvec}
 \eea
 with
 \be
 {\cal N}_v'=\cosh(i\pi\hat\k)\sin\pi(\frac{1}{2}+\hat
     m_v)-i\sinh(i\pi\hat\k)\cos\pi(\frac{1}{2}+\hat m_v).
 \ee
The conformal weight of the massive vector field is just
$h^v=\frac{1}{2}+\hat{m}_v$, where the identification
(\ref{identn}) has to be taken into account in $\hat{m}_v$. It is
easy to see that the correlator is consistent with the CFT
prediction (\ref{left}).

The quasi-normal modes could be read easily and are characterized
by exactly the same relation (\ref{quasiext}).

The cross section is just
 \be\label{sectionvec}
 \s \sim
 \frac{\G(2-2h^v)}{\pi\G(2h^v)}|\G(h^v-i\tilde{k})|^2
 \cos(h^v\pi)\sinh(i\pi\tilde{k}),
 \ee
which is very similar to the one of the scalar field.

In short, for the extremal spacelike warped AdS$_3$ black hole,
the retarded Green's function is well consistent with
(\ref{left}). Since the right-moving temperature is zero, it
behaves as a ``chiral" theory, similar to the null warped black
hole.

\section{Conclusions and discussions}

In this paper, we computed real-time correlators of the scalar and
vector operators from the warped AdS/CFT correspondence. We
discussed both the spacelike stretched and the null warped case.
In these cases, we can solve the equations of motion and obtain
the correlators analytically such that we can compare in a precise
way  the results got from gravity calculation with the CFT
prediction which are much restricted by 2D conformal invariance.
We found that the retarded correlators were in good match with the
prediction of dual CFT, up to a renormalization factor. This
strongly supports the warped AdS/CFT correspondence.

From the retarded correlators, we read out the scalar and vector
quasi-normal modes in the warped black hole background. They
correspond to the poles of the retarded Green's correlators in the
momentum space. The results are the same as the ones found in
\cite{ChenXu2}. Moreover, we obtained the cross section of the
scalar and vector field scattering the black holes from the
imaginary part of the retarded real-time correlators. Once again,
the cross sections are well consistent with the prediction of dual
CFT.

In our study, we noticed that the extremal spacelike warped black
hole behaves quite similarly to the null warped black hole. In
both cases, there exist only one temperature in dual CFT, which
seems suggest that the dual CFT is ``chiral". However, this
differs from the chiral gravity\cite{Li:2008dq} in which case
there does exist only one sector. We know that from the warped
AdS/CFT correspondence on the spacelike case, the dual CFT
actually has two independent sectors. The absence of one sector in
our study comes from the extremality of the black hole. This
indicates that in the null warped case, there may exist two
sectors with nonvanishing central charges, in support of the
diffeomorphism anomaly argument.

 For the global warped AdS$_3$
backgrounds,  the presence of the angular momentum quantum number
$\tilde{k}$ has very interesting physical implications. Firstly it
appears in the conformal weight of various kinds of the operators.
It may make the conformal weight not real, in which case there
exist superradiance in the black hole backgrounds. It is
remarkable that even in this case, the warped AdS/CFT
correspondence still make sense. This is very similar to the
situation in Kerr/CFT
correspondence\cite{Bredberg:2009pv,{Hartman:2009nz}}. Another
interesting implication of the angular momentum is that it induce
a charge with respect to a chemical potential in the right-moving
sector of dual CFT. This is important to set up the
correspondence.

It would be interesting to compare our study on the warped AdS
black hole with  Kerr/CFT
correspondence\cite{AndyWei,Bredberg:2009pv,{Hartman:2009nz}}. In
the latter case, one cannot read the real-time correlators
directly. Instead one can compute the cross section of the
scattering and compare with the CFT prediction. This is feasible
since the asymptotical geometry of Kerr black hole is flat and
there is no ambiguity in defining the ingoing and outcoming waves.
In practice, one has to divide the spacetime into several regions,
with one of them being the near horizon region, and study the
perturbations in different regions and try to glue them together.
Since the Kerr/CFT correspondence is actually the duality between
the quantum gravity on near horizon geometry of (near-)extremal
Kerr black hole (NHEK) with a 2D CFT, one has to figure out the
contribution at the NHEK region and compare them to CFT. While for
the warped AdS spacetime, it is hard to well define the ingoing
and outcoming waves  to get the absorption cross section, as shown
in the study of \cite{Oh:2008tc,{Kao:2009fh}}. Nevertheless, one
may bypass this obstacle by calculating directly the real-time
correlators from the prescription developed in usual AdS/CFT
correspondence. One can compare the Green's functions on both
sides. Furthermore, one may read the cross section  from the
real-time correlator easily, and find the agreement with CFT
prediction. Considering the fact that the near horizon geometry of
extremal Kerr black hole is a warped AdS spacetime, one would not
be surprised to find the similarity in the dictionaries of two
correspondences.

From our study, we have shown that both in real-time correlators
and absorption cross sections, the nontrivial dependence on Gamma
functions are in perfect match in warped AdS/CFT correspondence.
However, there is a discrepancy on normalization factor. From CFT
prediction, such a normalization factor should have nontrivial
temperature dependence, which is closely related to the conformal
weights of the operators. Our calculation failed to recover such
factors. It would be interesting to investigate this issue
further.

On the other hand, our investigation also suggests that the
prescription to get real-time correlator is still effective even
for the warped AdS spacetime which has a nontrivial conformal
boundary, even though it was developed in the usual AdS/CFT
correspondence. This may have profound implications in the recent
study of AdS/CMT correspondence. In fact, the null warped
backgrounds appear in the study of non-relativistic AdS/CFT of
cold
atoms\cite{Son:2008ye,{Balasubramanian:2008dm},{Herzog:2008wg},{Adams:2008wt},{Maldacena:2008wh}}
and also as the gravity dual of a Lifshitz field theory with
anisotropic
scaling\cite{Kachru:2008yh,{Balasubramanian:2009rx},{Horava:2009vy}}.
Our result gives strong support to apply the prescription in these
areas.

\section*{Acknowledgments}

 BC would like to thank Grey college, Durham University for
hospitality during his visit.
 The work was partially supported by NSFC Grant
No.10535060,10775002,10975005, and NKBRPC (No. 2006CB805905).

\ed


\end{thebibliography}

\end{document}